\documentclass[%
 reprint,
superscriptaddress,
 amsmath,amssymb,
 aps,
]{revtex4-2}

\usepackage{graphicx}
\usepackage{dcolumn}
\usepackage{bm}
\usepackage{multirow}
\usepackage[dvipsnames]{xcolor}

\begin{document}

\preprint{APS/123-QED}

\title{Theoretical description of optical and x-ray absorption spectra of MgO including many-body effects}

\author{Vijaya Begum}
\affiliation{Department of Physics and Center for Nanointegration Duisburg-Essen (CENIDE), University of Duisburg-Essen, Duisburg, Germany}

\author{Markus E. Gruner}
\affiliation{Department of Physics and Center for Nanointegration Duisburg-Essen (CENIDE), University of Duisburg-Essen, Duisburg, Germany}

\author{Christian Vorwerk}
\affiliation{Institute f\"ur Physik and IRIS Adlershof, Humboldt-Universit\"at zu Berlin, Berlin, Germany\\ and European Theoretical Spectroscopy Facility}

\author{Claudia Draxl}
\affiliation{Institute f\"ur Physik and IRIS Adlershof, Humboldt-Universit\"at zu Berlin, Berlin, Germany\\ and European Theoretical Spectroscopy Facility}

\author{Rossitza Pentcheva}
\affiliation{Department of Physics and Center for Nanointegration Duisburg-Essen (CENIDE), University of Duisburg-Essen, Duisburg, Germany}

\date{\today}

\begin{abstract}
   Here we report the optical and x-ray absorption (XAS) spectra of the wide-band-gap oxide MgO using density functional theory (DFT) and many-body perturbation theory
  (MBPT). Our comprehensive study of the electronic structure
  shows that while the band gap is underestimated with the exchange-correlation 
	functional PBEsol (4.58 eV) and the hybrid functional HSE06 (6.58 eV) compared
  to the experimental value (7.7 eV), it is significantly improved (7.52 eV)
  and even overcompensated (8.53 eV) when quasiparticle corrections are
  considered. Inclusion of excitonic effects by solving the Bethe-Salpeter equation (BSE) yields the optical spectrum in excellent agreement with experiment. Excellent agreement
  is observed also for the O and Mg K-edge absorption spectra, demonstrating the importance of the electron-hole interaction within MBPT. Projection of the electron-hole coupling coefficients
  from the BSE eigenvectors on the band structure allows us to determine the origin
  of prominent peaks and identify the orbital character of the relevant
    contributions. The real space projection of the lowest energy exciton wavefunction of
  the optical spectrum indicates a Wannier-Mott type, whereas the first
  exciton in the O K-edge is more localized. 
\end{abstract}
\maketitle


\section{\label{sec:Introduction}Introduction}
MgO is one of the most extensively studied oxides which is used as a 
substrate material and in various heterostructures with applications related to 
tunneling magnetoresistance \cite{Maruyama,Sofin,Yang}. As a wide band gap insulator with 
a measured optical band gap of 7.7 \cite{Roessler} from absolute-reflectance measurements with UV radiation
and 7.83 eV \cite{Whited} from thermoreflectance spectroscopy, 
this material is employed e.g. in transient x-ray spectroscopy and time-dependent density-functional theory 
(DFT) calculations aiming to unravel the propagation of excitations across the interface in
metal-insulator heterostructures \cite{Melnikov,Gruner,Rothenbach,Beyazit}. 
Understanding spectroscopic features from first-principles requires accurate modeling beyond
the ground state properties including excitations of different origin and energy scale.

The structural and electronic properties of MgO  have been widely studied with 
first-principles calculations \cite{Schleife-2006,Fuchs-2008,Shishkin}. DFT calculations with semilocal 
functionals yield a fundamental band gap of 4.88, 4.50 and 4.76 eV ~\cite{Wang,Schleife-2006,Shishkin}, respectively.
Many-body perturbation theory (MBPT) calculations employing Hedin's \textit{GW}
approximation \cite{Hedin} render an increased fundamental gap of 6.8 and 7.25 eV \cite{Fuchs-2008,Shishkin}, 
respectively, which is still lower than the experimental one. 

\noindent The optical spectrum, calculated by Wang \textit{et al.} \cite{Wang} using the 
local density approximation (LDA) as the exchange-correlation functional for the DFT calculation and
subsequently including \textit{GW} and excitonic corrections agrees 
with experiment \cite{Palik} w.r.t. peak positions up to 12 eV 
whereas  the amplitude of the peaks beyond the first one is overestimated
due to the limited number of unoccupied bands employed in the BSE corrections. Schleife
\textit{et al.} \cite{Schleife-2006} studied the frequency-dependent dielectric
function for different MgO polymorphs -- wurzite, zinc blende, and rocksalt -- in
the independent particle (IP) approximation using the generalized gradient approximation in the PW91 
parametrization \cite{PW91}. Good agreement with experiment concerning the peak positions was obtained 
by including excitonic corrections with BSE, based on the Kohn-Sham (KS) eigenenergies and a scissors 
operator to describe the QP eigenenergies \cite{Schleife-2009,Schleife-2012}.

While optical spectroscopy probes excitations from valence bands, x-ray
absorption spectroscopy (XAS) probes those from the strongly localized core
states. A common approach to model XAS is the final state rule (FSR) \cite{Barth} based on
Fermi's Golden rule, where the effects of screening of the
core-hole (the so-called \textit{final-state effects}) are
calculated in a supercell. Alternatively, XAS can be described by considering
quasiparticle and excitonic effects within MBPT by using $GW$
and solving the BSE. Rehr \textit{et al.} \cite{Rehr} showed that
while both approaches led to similar overall features in the O
and Mg K-edge spectra of MgO, BSE calculations result in
better agreement with experiment at high transition energy due to the
non-local treatment of the exchange interaction. Recent implementations of BSE in
all-electron codes \cite{Laskowski,Vorwerk-2017} with explicit treatment of core
states have demonstrated very good agreement with experiment for
the XAS spectra of TiO${_2}$ (rutile and anatase), PbI${_2}$, and
CaO \cite{Vorwerk-2017}. The latter approach is adopted in this work.

Here we describe both the optical and x-ray absorption spectra of bulk MgO
including many-body effects. As a first step, we perform the $G{_0}W{_0}$
corrections starting from Kohn-Sham (KS) wavefunctions. We show that careful consideration of the electron-hole interaction with BSE is essential to achieve
agreement with experiment for both valence and core excitation spectra. In particular, the optical 
spectrum calculated with two different DFT functionals (PBEsol and HSE06) including the $G{_0}W{_0}$ 
and BSE corrections are consistent with experiment~\cite{Roessler,Bortz} and previous theoretical work \cite{Schleife-2009,Schleife-2012} and yield an improved agreement regarding the intensity of the peaks at higher energies, highlighting the importance of quasiparticle and excitonic effects. 

Previous studies have shown that a dense \textbf{k}-mesh is
required for the sampling of the Brillouin zone to describe sufficiently the
localization of the excitonic wave function and the fine structure in the
vicinity of the absorption edge \cite{Fuchs-2008}. Here, we use a model for the 
static screening with parameters fitted to the $G{_0}W{_0}$ calculation, to solve the BSE 
(so-called model BSE \cite{Fuchs-2008,Liu}) starting directly from DFT wavefunctions on a denser \textbf{k}-mesh, 
which improves in particular the low energy range (7$-$11 eV). The results for the model BSE are  
presented in Appendix \ref{sec:mBSE}. Beyond previous work we provide a 
thorough analysis of interband transitions contributing to the peaks in the optical spectrum. Further 
insight into the nature of the first bound exciton is given by the real-space visualization of its 
wave function.

Employing the \texttt{exciting} code, the O and Mg K-edge XAS spectra calculated with BSE show very good agreement with the experimental spectra \cite{Luches} and with previous theoretical results using the FSR \cite{Rehr}. Knowledge of the origin of peaks is essential for the interpretation of x-ray spectra. The main incentive of this study is to identify the nature of
transitions which contribute to the peaks and analyze the character of the first
exciton in the O K-edge both in real and reciprocal space.%

The paper is structured as follows: the details of the calculations are
presented in Section \ref{sec:CompDetail}, followed by the discussion of the
results in Section \ref{sec:Result}. We start with the electronic properties of
MgO in \ref{subsec:properties} and then  compare the optical spectra calculated
with two different starting exchange-correlation functionals in
\ref{subsec:Optprop}. Subsequently, we analyze the transitions in reciprocal
space to derive the origin of contributions to the peaks in the spectrum. In
subsection \ref{sec:XASprop}, we present the XAS spectra of the O and Mg K-edge
and identify the underlying transitions in reciprocal space for the prominent
peaks. Finally, subsection \ref{sec:RealspaceProj} is
dedicated to the real-space visualization of the first exciton of the optical
and the O K-edge x-ray absorption spectrum. The results are summarized in
Section \ref{sec:Summary}, followed by two appendices showing a comparison of the optical spectra obtained with VASP and \texttt{exciting} and the optical spectrum with the model BSE.

\begin{figure*}[!htp]
\includegraphics[width=1.0\textwidth]{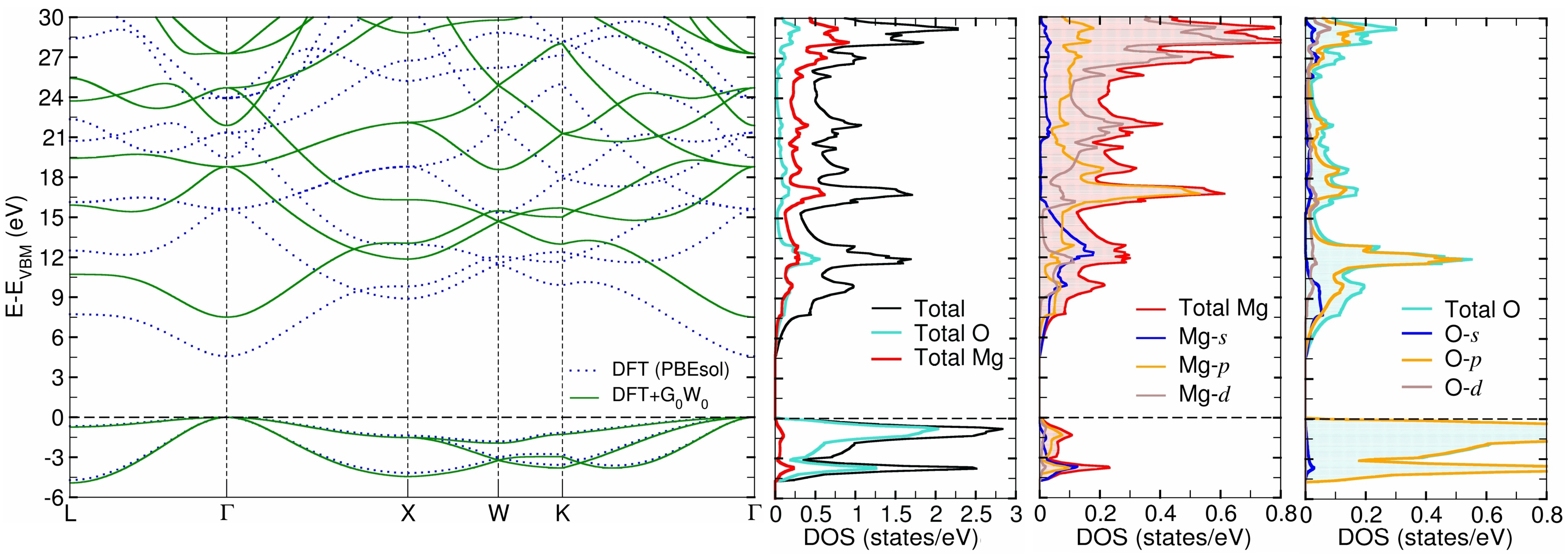}%
\caption{\label{fig:Bandstr-PDOS}(a) Kohn-Sham and $G{_0}W{_0}$ band structure and (b-d) total and projected density of states (PDOS) of MgO calculated with PBEsol within VASP.}
\end{figure*}
\section{\label{sec:CompDetail}Computational details}
The DFT calculations are performed with the VASP code (version
5.4.4) \cite{Vasp-96,Vasp2-96}, using pseudopotentials in combination with the
projector augmented wave (PAW) method \cite{Vasp3-99}, and the \texttt{exciting} code \cite{Gulans-2014} (version Nitrogen) employing the all-electron full-potential (linearized) augmented planewave $+$ local orbital [(L)APW$+$lo] method. For the exchange-correlation
functional we chose the generalized gradient approximation (GGA) in the
implementation of Perdew, Burke, and Ernzerhof (PBE96) \cite{PBE-96}, PBEsol
\cite{PBEsol-08,PBEsol-09}, and the hybrid functional, HSE06 
\cite{HSE-03,HSE-05}. The equilibrium lattice constant determined with the different functionals amounts to 4.24~\AA\ (PBE96), 4.21~\AA\ (PBEsol), and 4.20~\AA\ (HSE06), the experimental one being 4.212~\AA\ \cite{Landolt}.

For the calculation of the optical spectrum with VASP, we
have performed single-shot $G{_0}W{_0}$ on top of the KS wavefunctions obtained
with two DFT functionals, PBEsol and HSE06 and subsequently included excitonic
corrections by solving the BSE. For all the BSE calculations
the Tamm-Dancoff-approximation (TDA) \cite{Dancoff} is adopted. The calculations
are performed for a two-atom unit cell with a $\Gamma$-centered
15$\times$15$\times$15 \textbf{k}-mesh (unless otherwise specified) with a
plane-wave cut-off energy of 650 eV. $GW$ PAW pseudopotentials for excited
properties were employed in all the calculations with two valence electrons for
Mg: $3s^2$ and six for O: $2s^2$, $2p^4$. 192 unoccupied bands are used for
both the DFT and single-shot $G{_0}W{_0}$ calculations with 100 frequency-grid
points. For the optical spectrum a Lorentzian broadening of 0.3 eV is used.

Employing PBEsol~\cite{PBEsol-08,PBEsol-09} as the starting exchange-correlation functional for the 
ground-state calculation, single-shot $G{_0}W{_0}$ calculations are also performed with the
\texttt{exciting} code \cite{Gulans-2014} together with BSE \cite{Sagmeister-2009} (within TDA) for
the optical and x-ray absorption spectra \cite{Vorwerk-2017}. 
A $\Gamma$-centered 11$\times$11$\times$11 mesh shifted by (0.09, 0.02,0.04) is
employed for the calculations. Muffin-tin radii of 1.058 and 0.767~\AA\ for
Mg and O, respectively, are used with a basis set cut-off $R_{MT}|\mathbf{G}+\mathbf{k}|_{max}=7$, 
and the lattice constant is set to the PBEsol value of
4.21~\AA . The energy threshold to include the local field effects in the excited
properties, $|\mathbf{G}+\mathbf{q}|_{max}$, is set to 4.5 a.u.$^{-1}$ for the
optical and O K-edge, and 1.5 a.u.$^{-1}$ for the Mg K-edge absorption spectra. 
The exchange-correlation functional PBEsol is
employed for the Kohn-Sham (KS) states and a total of 192 unoccupied bands are
considered in the ground state and $G{_0}W{_0}$ calculation for the optical and
O and Mg K-edge x-ray absorption spectra. For the
optical spectrum in the BSE calculation, four occupied and five unoccupied bands are considered, while
eight unoccupied bands were taken into account for the XAS spectra.  A Lorentzian broadening with a 
width of 0.55 eV is applied to the spectra to mimic the
excitation lifetime. The atomic structures and isosurfaces are visualized with
the VESTA software \cite{Vesta} and the band structure is calculated with the
Wannier90 \cite{Wannier90} package in VASP.

\section{\label{sec:Result}Results}
\subsection{\label{subsec:properties}Electronic properties}
We start our analysis by comparing the electronic properties obtained from DFT
calculations with three different functionals, namely PBE96, PBEsol and HSE06.
Table \ref{tab:ExcGWbg} presents the band gap calculated with VASP. With PBE96
(4.49 eV) and  PBEsol (4.58 eV), the band gaps are considerably underestimated,
consistent with previous calculations \cite{Schleife-2006,Fuchs-2008,Shishkin}.
On the other hand, HSE06 renders a band gap of 6.58 eV closest but still below
the experimental value of 7.7 and 7.83 eV \cite{Roessler,Whited}.
The $G{_0}W{_0}$ band gap obtained with PBEsol (7.52 eV) is closest to experiment, whereas 
a somewhat lower value (7.26 eV) is obtained with PBE96 which is in agreement with 
the value of 7.25 eV from Ref.~\cite{Shishkin}. 
The latter study ~\cite{Shishkin} has also addressed the effect of self-consistent quasiparticle
correction cycle on the optical properties: self-consistency in $G$ while keeping $W_0$ constant ($GW_{0}$) 
increased the band gap to 7.72 eV, while fully self-consistent $GW$ led to an overestimated band gap of 8.47 eV 
and was attributed to the missing vertex corrections in the self-consistency cycle. While the size 
of the band gap may be reproduced by considering (partial) self-consistency in $GW$, our results show that 
the inclusion of excitonic effects (see Sections \ref{subsec:Optprop} and \ref{sec:XASprop}) is essential 
in order to describe the relevant features and the shape of the spectrum. We note that the 
$G{_0}W{_0}$ band gap with the hybrid HSE06 functional is also overcorrected (8.53 eV). This is consistent 
with previous findings that the effects of the starting exchange-correlation functional are large at the 
independent-particle level, but the differences are reduced when considering quasiparticle~\citep{Jiang} and 
eventually excitonic effects~\cite{Begum}.

\begin{figure*}[!htp]
\includegraphics[width=0.85\textwidth]{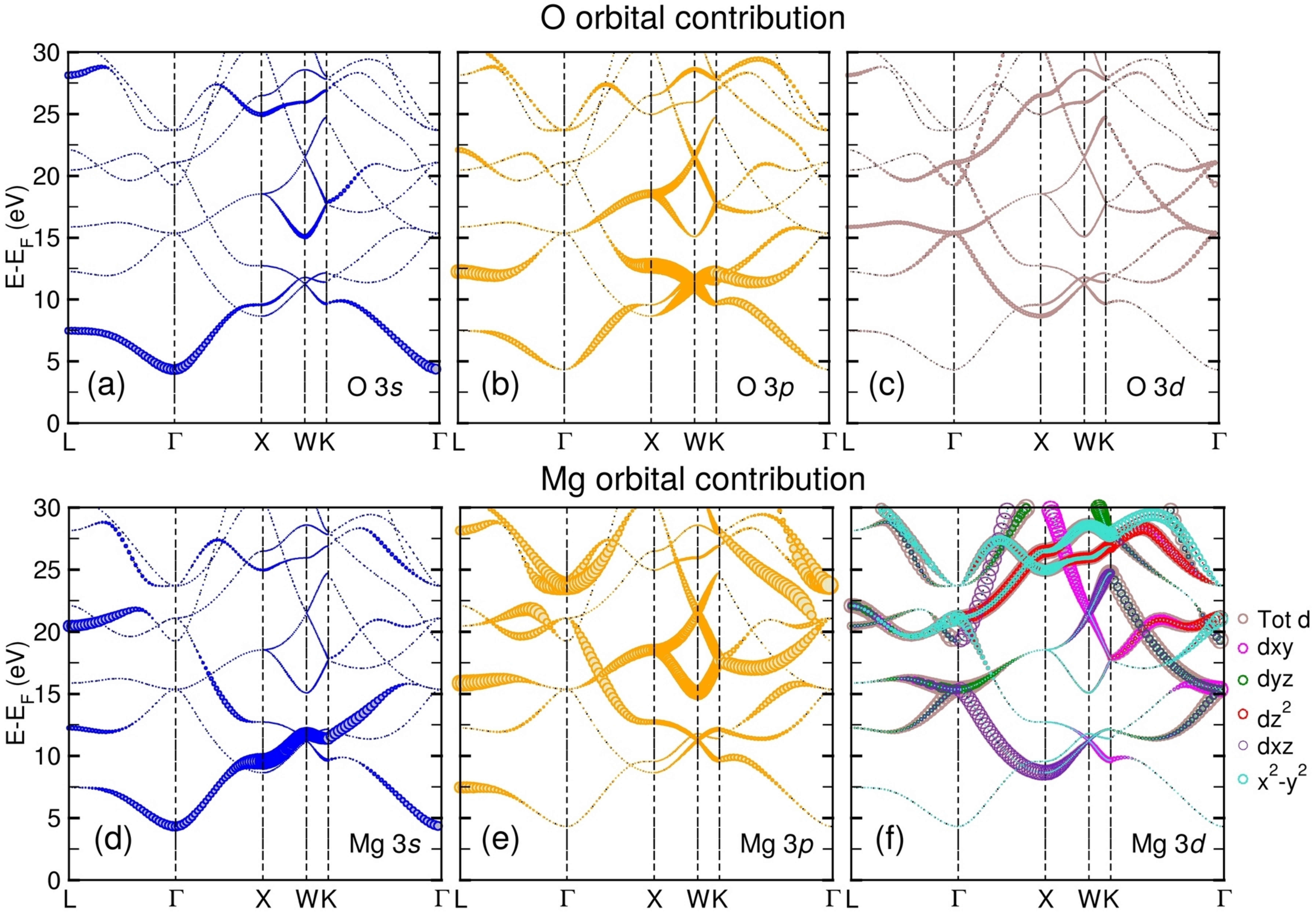}%
\caption{\label{fig:Vasp-Orbital}Oxygen (a-c) and Mg (d-f) orbital-resolved
  contributions projected on the ground state band structure within VASP.}
\end{figure*}

Since PBEsol and HSE06 provide better electronic properties as
compared to PBE96 we continue the analysis with those. In Fig.~\ref{fig:Bandstr-PDOS}a 
the Kohn-Sham and $G{_0}W{_0}$ band structure with the PBEsol functional is plotted 
along high-symmetry points, showing a direct ($\Gamma-\Gamma$) band gap. The inclusion of 
quasiparticle effects in the $G{_0}W{_0}$ calculation leads to a nearly rigid shift of the unoccupied
Kohn-Sham bands to higher energies. The top of the valence band (VB)
consists mainly of O $2p$ states (cf. the projected density of states in
Figs. \ref{fig:Bandstr-PDOS}b-d) with low dispersion along the
$L-\Gamma-K$ direction, whereas the lower bands are more dispersive.
Further insight into the orbital-resolved contributions of O
and Mg on the band structure is provided in Fig.~\ref{fig:Vasp-Orbital}. The
bottom of the conduction band (CB) comprises hybridized O $3s$,
$3p$, and Mg $3s$ states that are highly dispersive along the
$L-\Gamma-X$ and $K-\Gamma$ directions (cf. Figs.
\ref{fig:Bandstr-PDOS}c, d and Figs. \ref{fig:Vasp-Orbital}a, b, and d). In the
range of 4.5 - 11 eV beyond the CB minimum, $3s$, $3p$, and Mg $3s$ states prevail, 
whereas above 11 eV O $3p$ states become
predominant, followed by  Mg $3p$ and $3d$ states above 15 eV
(cf. Fig.  \ref{fig:Bandstr-PDOS}d and Figs. \ref{fig:Vasp-Orbital}e, f). 
We will further analyze the ion- and orbital projections in the band structure in 
Section \ref{subsubsec:exciton-opt} and Section \ref{sec:XASprop} to correlate the 
contributions with the optical and XAS spectra.

\begin{table}[!htp]
\caption{\label{tab:ExcGWbg} Comparison of the fundamental band gap from the  DFT and the
  $G_0W_0$ calculation with different starting functionals.}
\begin{ruledtabular}
\begin{tabular}{lcccc}
\textrm{}&
\textrm{E$_{xc}$}&
\textrm{DFT}&
\textrm{$G_{0}W_{0}$}&
\textrm{Experiment}\\
\colrule
\multirow{3}{4em}{$E_g$ (eV) ($\Gamma-\Gamma$)} & PBE96 & 4.49 & 7.26 & \multirow{3}{4em}{7.7\footnotemark[1], 7.83\footnotemark[2]}\\
 & PBEsol & 4.58 & 7.52 &\\
 & HSE06 & 6.58 & 8.53 &\\
\end{tabular}
\end{ruledtabular}
\footnotetext[1]{Reference \citenum{Roessler}}
\footnotetext[2]{Reference \citenum{Whited}}
\end{table}

\subsection{\label{subsec:Optprop}Optical properties}
Starting from the electronic structure presented in the last
section, we determine the optical spectrum including also many-body effects. We discuss the
effect of approximations to the exchange-correlation functional, namely PBEsol and
HSE06, on the spectra and the role of inclusion of $G_{0}W_{0}$ and excitonic
corrections with BSE. In addition, the interband transitions
responsible for the spectral features are analyzed in reciprocal
space.
\subsubsection{Optical spectrum within IP approximation and inclusion of $G_0W_0$ corrections} 
\begin{figure*}[!htp]
\includegraphics[width=0.98\textwidth]{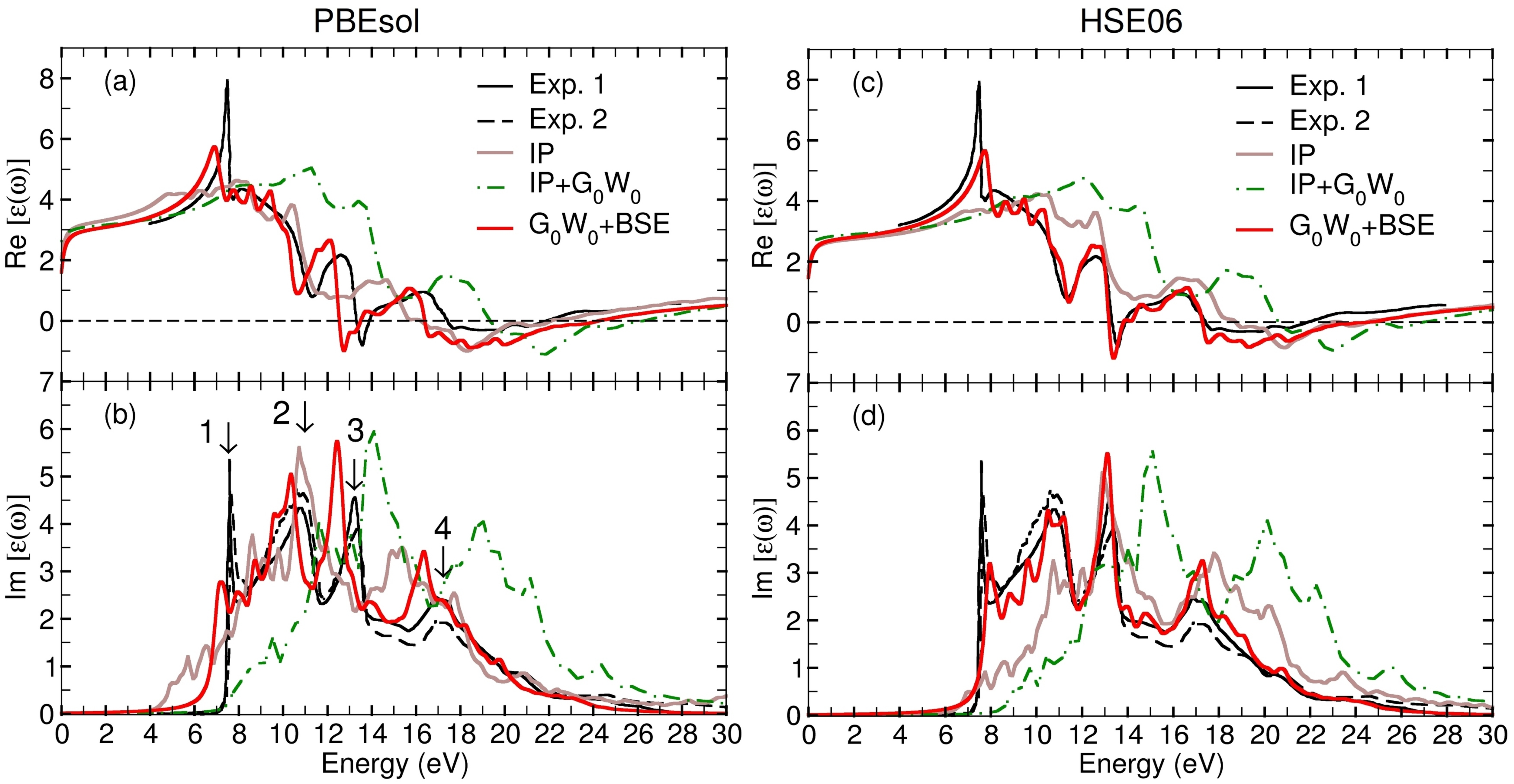}
\caption{\label{fig:optics-vasp} Optical spectrum of bulk MgO obtained with
  VASP: (a), (c) real part and (b), (d) imaginary part of the dielectric
  function for PBEsol and HSE06 as the starting functional, respectively. A
  Lorentzian broadening of 0.3 eV is employed for all the calculated spectra.
  The IP, IP$+G_{0}W_{0}$, and $G_{0}W_{0}+$BSE results are shown by brown solid,
  green dash-dotted, and red solid lines, respectively. Additionally, the
  experimental data from Exp. 1 \cite{Roessler} (black solid line), and Exp. 2 \cite{Bortz} (black dashed line) are displayed.}
\end{figure*}

The calculated optical spectra are plotted in Fig.~\ref{fig:optics-vasp}
together with the experimental ones~\cite{Roessler,Bortz}. The imaginary part of
the experimental dielectric function shows four prominent peaks (marked in Fig.
\ref{fig:optics-vasp}b): the first two at $\sim$7.7 eV and 10.7 eV are of nearly
equal intensity, the third and fourth peak are at 13.32 and 16.9 eV,
respectively, with a smaller intensity of the latter. We start our analysis by
considering the results from the independent particle (IP)
approximation using the KS eigenvalues calculated with the functionals PBEsol
and HSE06. The imaginary part of the dielectric function has its
onset at 4.58 and 6.58 eV for PBEsol and HSE06, respectively, below the
experimental spectrum, due to the underestimation of the band gap (cf. Table \ref{tab:ExcGWbg}). 
Moreover, prominent peaks in the imaginary part of the spectrum are observed at $\sim$8.5,
11, and 15.5 eV for PBEsol and at around 11, 13, and 17.5 eV for HSE06,
corresponding to pronounced band transitions that coincide with points of
inflection in the real part of the spectrum.

Inclusion of many-body effects within the $G_0W_0$ approximation results in a
blue shift by $\sim$3 eV (PBEsol) and 2 eV (HSE06) compared to the IP spectra,
due to the increased band gaps within $G_0W_0$. This strong effect is attributed to the 
weak dielectric screening in MgO \cite{Wang}. In Figs.~\ref{fig:optics-vasp}b, d sharper
features emerge in $\epsilon_2$ with peaks at $\sim$11.8, 14, and 19 eV (PBEsol),
that are shifted to higher energies at $\sim$ 12.5, 15, and 20.5 eV (HSE06).
The real part of the spectrum in Figs.~\ref{fig:optics-vasp}a, c exhibits
weaker/smoothened features compared to experiment \cite{Roessler} (Fig.
\ref{fig:optics-vasp}b, d). The macroscopic static electronic dielectric constant,
$\epsilon_{\infty}=$Re $\epsilon(\omega=0)$ obtained with PBEsol and HSE06 is
presented in Table \ref{tab:Exc-eps}. Within the IP approximation,
$\epsilon_{\infty}$ is overestimated for PBEsol (3.29) compared to the
experimental value 2.94~\cite{Roessler}, similar to previous results with
GGA-PW91 (3.16) \cite{Schleife-2009}. We also included the local field effects in the 
IP calculation and find that the dielectric constant decreases from 3.29 to 3.04 (PBEsol) 
and 2.76 to 2.57 (HSE06). A similar trend was observed in the work of 
Gajdoš \textit{et al.}~\cite{Gajdos-2006} 
for semiconductors as Si, SiC, AlP, GaAs, and for insulator as diamond (C).
On the other hand, with the hybrid functional, $\epsilon_{\infty}$ is underestimated (2.76). Upon including
quasiparticle effects ($G_0W_0$), the values are substantially reduced to 2.78
(PBEsol) and 2.53 (HSE06). 

\begin{table}[!htp]
\caption{\label{tab:Exc-eps} Comparison of the macroscopic static electronic
  dielectric constant $\epsilon_{\infty}$ in the IP approximation and after
  $G_0W_0$ and BSE with different DFT functionals.} 
\begin{ruledtabular}
\begin{tabular}{lcccc}
\textrm{E$_{xc}$}&
\textrm{IP}&
\textrm{G$_{0}$W$_{0}$}&
\textrm{BSE}&
\textrm{Experiment}\\
\colrule
PBEsol & 3.29 & 2.78 & 3.08 & \multirow{3}{4em}{2.94\cite{Roessler}}\\
 HSE06 & 2.76 & 2.53 & 2.81 &\\
\end{tabular}
\end{ruledtabular}
\end{table}
\subsubsection{Optical spectrum with excitonic corrections\label{subsec:GW+BSW}}
Additional to the quasiparticle corrections, we consider the
effects arising from electron-hole interaction by solving the Bethe-Salpeter
equation. The calculations are performed with four occupied and five unoccupied
bands which are sufficient to evaluate the optical spectrum up to 30 eV.

The inclusion of excitonic effects leads to a redistribution of the spectral
weight to lower energies w.r.t. the $G_0W_0$ spectrum and the emergence of a
sharp peak at the absorption onset. With PBEsol as the starting point in Fig.
\ref{fig:optics-vasp}b, the agreement with experiment w.r.t. spectral shape is improved, but the onset
of the imaginary part of the dielectric function is $\sim$0.7 eV lower than experiment \citep{Roessler,Bortz}. 
The prominent peaks are at $\sim$7.0 eV, 10 eV, 12.4 eV, and 16 eV. 
In the real part of the spectrum, the sign reversal at 12.8 eV indicates a plasmonic resonance.

On the other hand, using HSE06 as the starting functional, the real
and imaginary part of the spectrum are in excellent agreement with experiment.
The peak positions of the distinct features and
the plasmonic resonance at 13.4 eV coincide with the experimental ones
\cite{Bortz}. The four peaks of $\epsilon_2$ at
$\sim$8, 10.5, 13, and 17 eV are largely aligned with experiment, as shown
in Figs.~\ref{fig:optics-vasp}c and d. Further analysis of the origin of the
peaks in reciprocal space and the real-space projection of the first exciton are
provided in Section \ref{subsubsec:exciton-opt} and Section
\ref{sec:RealspaceProj}, respectively.  
The improved description w.r.t. the energetic positions and, to a lesser
extent, intensity of the peaks can be attributed to the description of the 
ground state with a hybrid functional HSE06, the larger number of unoccupied bands considered in the BSE
calculation and to performing BSE on top of $G_{0}W_{0}$. Furthermore, the
agreement to experiment concerning the macroscopic static electronic dielectric
constant $\epsilon_{\infty}$ is improved after BSE to 3.08 (PBEsol) and 2.81
(HSE06), respectively (cf. Table \ref{tab:Exc-eps}), also consistent with a 
previous value of 3.12 \cite{Schleife-2009}, where excitonic
corrections were included using the KS eigenenergies and a scissor shift
approach. We note that increasing the number of unoccupied bands to 9, leads to
slightly higher values for $\epsilon_{\infty}$ 3.16 (PBEsol) and 2.90 (HSE06),
the latter being in excellent agreement with experiment. In particular, more empty states are necessary for the calculation of the real part of the dielectric function from the Kramers-Kronig relation. Due to the high computational cost and an enhanced memory demand with more unoccupied bands, we proceed with five unoccupied bands for the further analysis.

\begin{figure*}[!htp]
\includegraphics[width=0.98\textwidth]{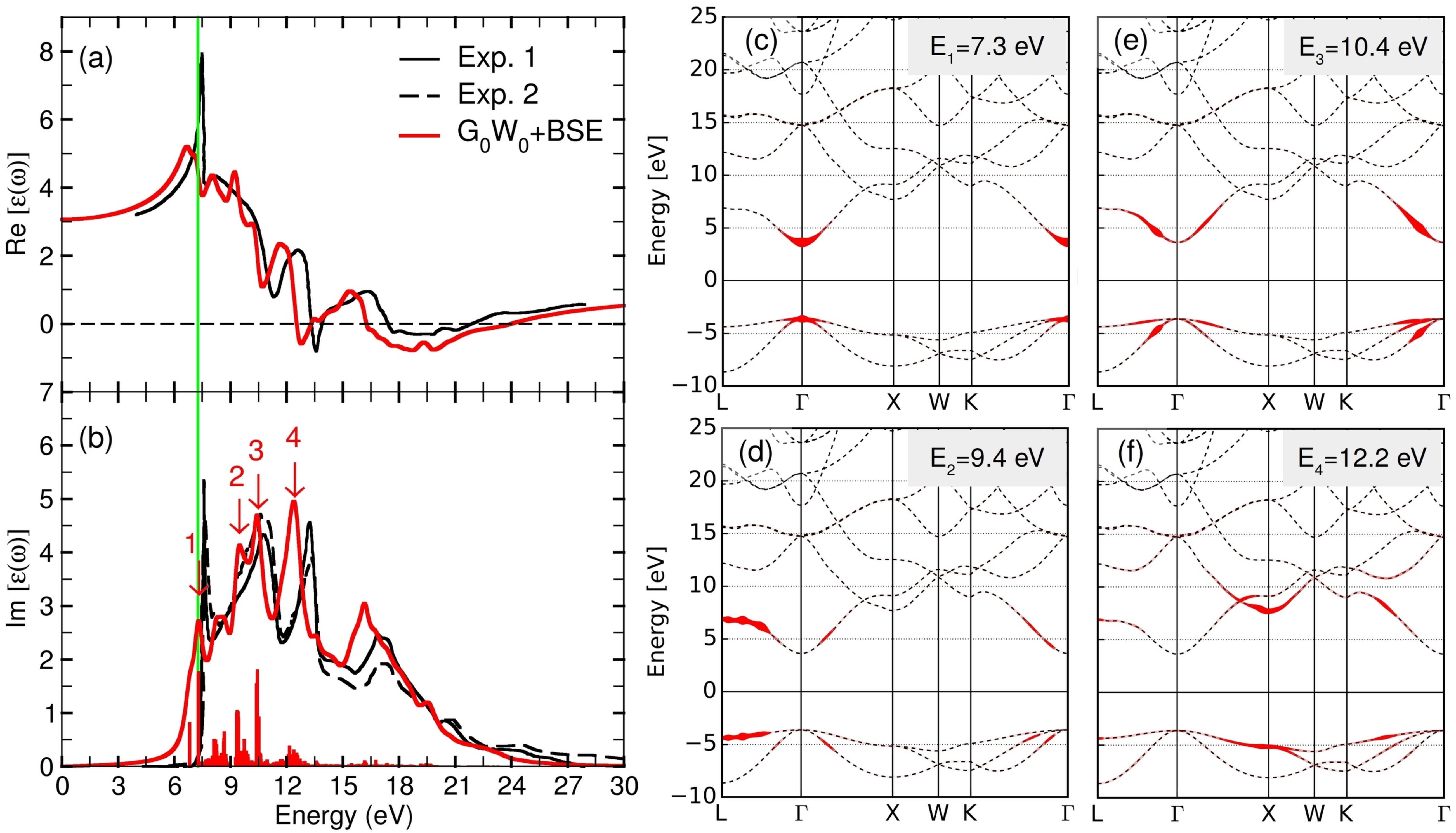}
\caption{\label{fig:optics-exc} Optical spectrum with PBEsol including many-body
  corrections calculated with the \texttt{exciting} code: (a) real and (b)
  imaginary part of the dielectric function. A Lorentzian broadening 0.3 eV is
  employed for the calculated spectrum for the $G_{0}W_{0}$+BSE corrections (red
  line) and the red vertical bars represent the oscillator strength (arb. units). 
  The direct band gap at 7.26 eV is marked by a vertical green line. Experimental       
  spectra from Roelssler \textit{et al.}~\cite{Roessler} (black solid line) and Bortz
  \textit{et al.}~\cite{Bortz} (black dashed line) are shown for comparison. (c$-$f) 
  electron-hole coupling coefficients represented as circles in reciprocal space 
  for the peaks at different energies marked in (b), where the size of the circle is proportional 
  to the magnitude of the e-h contribution.}
\end{figure*}

Furthermore, for the binding energy of the first exciton we obtain 
442 meV with PBEsol and 596 meV with HSE06.  A similar value of 429 meV was
obtained by Fuchs \textit{et al.} \cite{Fuchs-2008} employing the KS
eigenenergies (GGA) with a scissor shift of 2.98 eV and subsequently including
excitonic corrections. The overestimation of the binding energy w.r.t experiment (80 meV
\cite{Roessler}) may be
attributed to the fact that the ionic contributions to the static screening is
not considered \cite{Shindo,Zimmermann}.

\subsubsection{\label{subsubsec:exciton-opt}Analysis of spectral features in
reciprocal space} 

In order to identify the origin of the most prominent peaks,
we have performed calculations with the all-electron \texttt{exciting} code. The
real and imaginary part of the dielectric function for the $G_0W_0+$BSE with
PBEsol as the DFT functional and similar parameters (four occupied and five
unoccupied bands) are plotted in Figs. \ref{fig:optics-exc}a, b, and show good
agreement with experiment as well as the VASP result w.r.t. the
energetic positions of the peaks (a comparison of the spectra obtained with the two codes is provided in 
Appendix A and Fig. \ref{fig:Comparison-OptSpec}). The PBEsol band gap is 4.60 eV at the KS level and increases 
to 7.25 eV after quasiparticle corrections with $G_0W_0$ are included.
The most prominent peaks are marked in Fig.~\ref{fig:optics-exc}b and the corresponding e-h contributions 
are studied in Figs.~\ref{fig:optics-exc}c-f. We recall that the Bethe-Salpeter equation represents an
eigenvalue problem for an effective two-particle Hamiltonian~\cite{Rohlfing,Sagmeister-2009}:
\begin{equation}\label{eqn:BSE_EVP}
\sum_{v'c'\textbf{k}'}H_{vc\textbf{k},v'c'\textbf{k}'}A^{\lambda}_{v'c'\textbf{k}'}=E^{\lambda}A^{\lambda}_{vc\textbf{k}}.
\end{equation}
where E$^\lambda$ are the transition energies and
A$^\lambda_{vc\textbf{k}}$ are the corresponding states in
terms of $v \mathbf{k} \rightarrow c \mathbf{k}$ transitions. 
The e-h coupling coefficients for a particular transition displayed as circles
in Figs. \ref{fig:optics-exc}c-f are calculated from the BSE eigenvector
A$^\lambda$ as:
\begin{equation}
w^\lambda_{c\textbf{k}}=\sum_{c}\mid A^{\lambda}_{vc\textbf{k}} \mid^2, \, 
w^\lambda_{v\textbf{k}}=\sum_{v}\mid A^{\lambda}_{vc\textbf{k}} \mid^2.
\end{equation}

\begin{figure*}[!htp]
\includegraphics[width=1.0\textwidth]{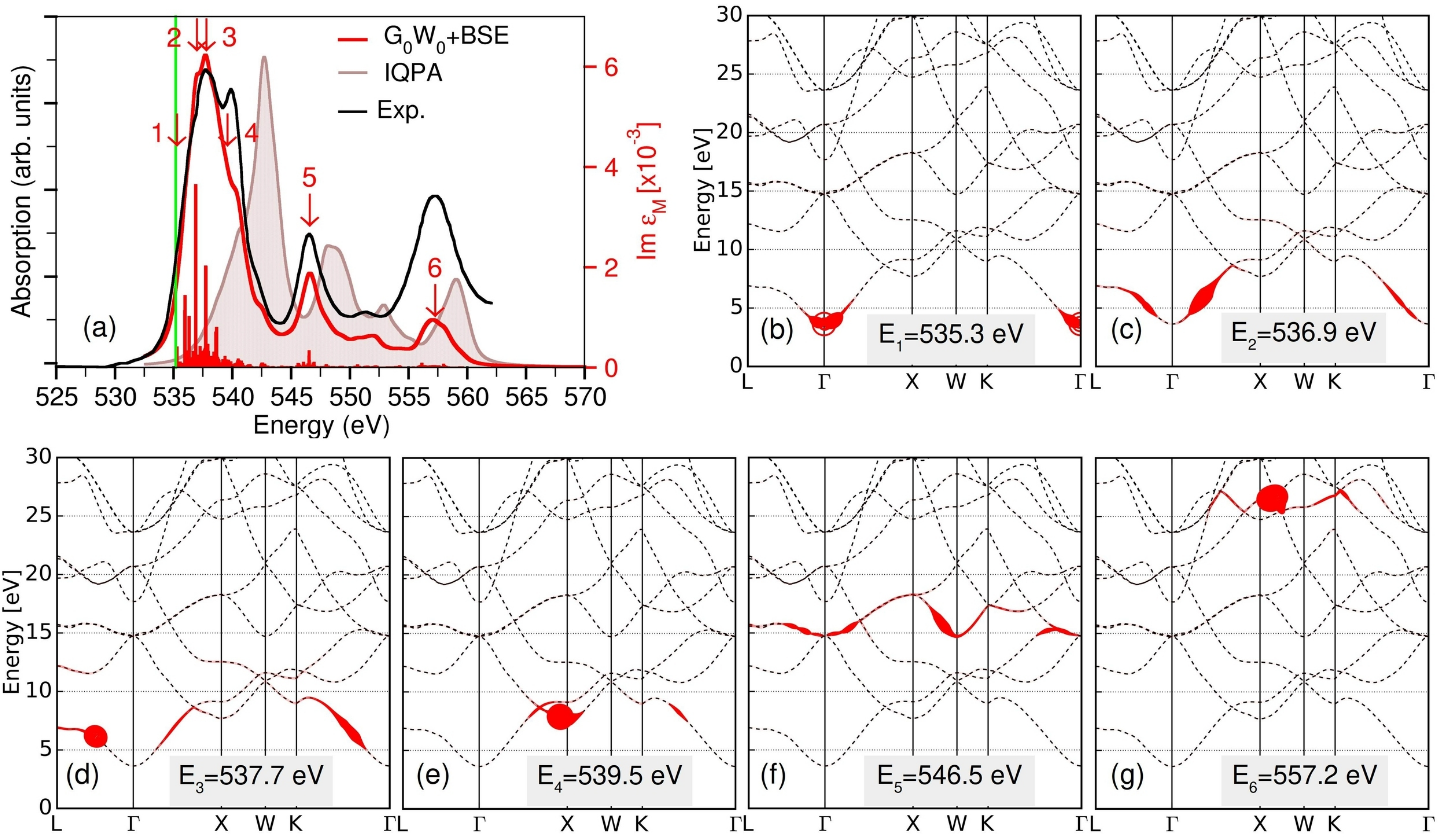}%
\caption{\label{fig:xas-O}XAS spectrum of the O K-edge using $G_{0}W_{0}$+BSE
  calculated with the \texttt{exciting} code: (a) calculated absorption spectra
  with $G_{0}W_{0}$+BSE (red line) and within the independent quasiparticle approximation 
  (IQPA, brown shaded area) are compared with experimental spectra from 
  Luches \textit{et al.}~\cite{Luches} (black line). A
  shift of 34.4 eV was applied to the calculated spectra to align to the first
  peak of the experiment and a Lorentzian broadening of 0.55 eV is adopted to
  mimic the excitation lifetime. The green line  at 535.2 eV marks the direct band gap. The red vertical bars represent the oscillator strength (arb. units). (b-g) excitonic contributions to the final states in the CB of the peaks marked in (a).}
\end{figure*}

The first exciton at 6.82 eV has a binding energy of 435 meV, close to the value obtained with VASP, 
as discussed in the previous section. This bound exciton contributes to the shoulder at the onset of 
the spectrum. The interband transitions responsible for the exciton and its real-space distribution 
are discussed in detail in Section \ref{sec:RealspaceProj}. In Fig. \ref{fig:optics-exc} a, the green line 
marks the fundamental band gap, below which the bound excitons lie.

The first peak at 7.3 eV (cf. Fig.~\ref{fig:optics-exc}c) arises due to
transitions from the top of the valence band (VB) to the bottom of the
conduction band (CB) around the $\Gamma$-point in reciprocal space. 
A comparison with the site and orbital- projected DOS (Fig.~\ref{fig:Bandstr-PDOS}) and band structure 
(Fig. \ref{fig:Vasp-Orbital}) reveals a mixed O $3s$, $3p$, and Mg $3s$ character. The second peak at 
9.4 eV involves interband transitions from the topmost VB to the lowest CB along $L-\Gamma-X$ and $\Gamma-K$. 
The CB is more dispersive along $L-\Gamma$ and has mixed O $3s$ and 
$3p$ character with Mg $3p$ contributions near $L$. The next peak at 10.4 eV
stems from transitions to the CB from deeper-lying valence bands along $L-\Gamma-X$
and $\Gamma-K$. The final peak at 12.2 eV, plotted in Fig.~\ref{fig:optics-exc}f,
results from transitions from the top of the VB to the higher-lying CB around $X$
as well as along $K-\Gamma$. In this energy range, the CB consists of O $3p$ and
Mg $3s$ and $3d_{xy}$,$d_{xz}$ states along $\Gamma-X$ and $K-\Gamma$. 

The influence of lattice screening on the optical spectrum is a topic of current research and so far 
there is no established framework to treat the exciton-phonon coupling that renormalizes the absorption 
spectra. Such effects should be assessed in future work.

\subsection{\label{sec:XASprop}X-ray absorption spectra}
We now turn to the x-ray absorption spectra of the O and Mg K-edge of bulk MgO
calculated with the \texttt{exciting} code. The ground state calculations were performed with the PBEsol 
exchange-correlation functional and the quasiparticle corrections were included with the single-step $G_0W_0$. 
Finally, the excitonic corrections were accounted for by solving BSE.
\begin{figure*}[!htp]
\includegraphics[width=1.0\textwidth]{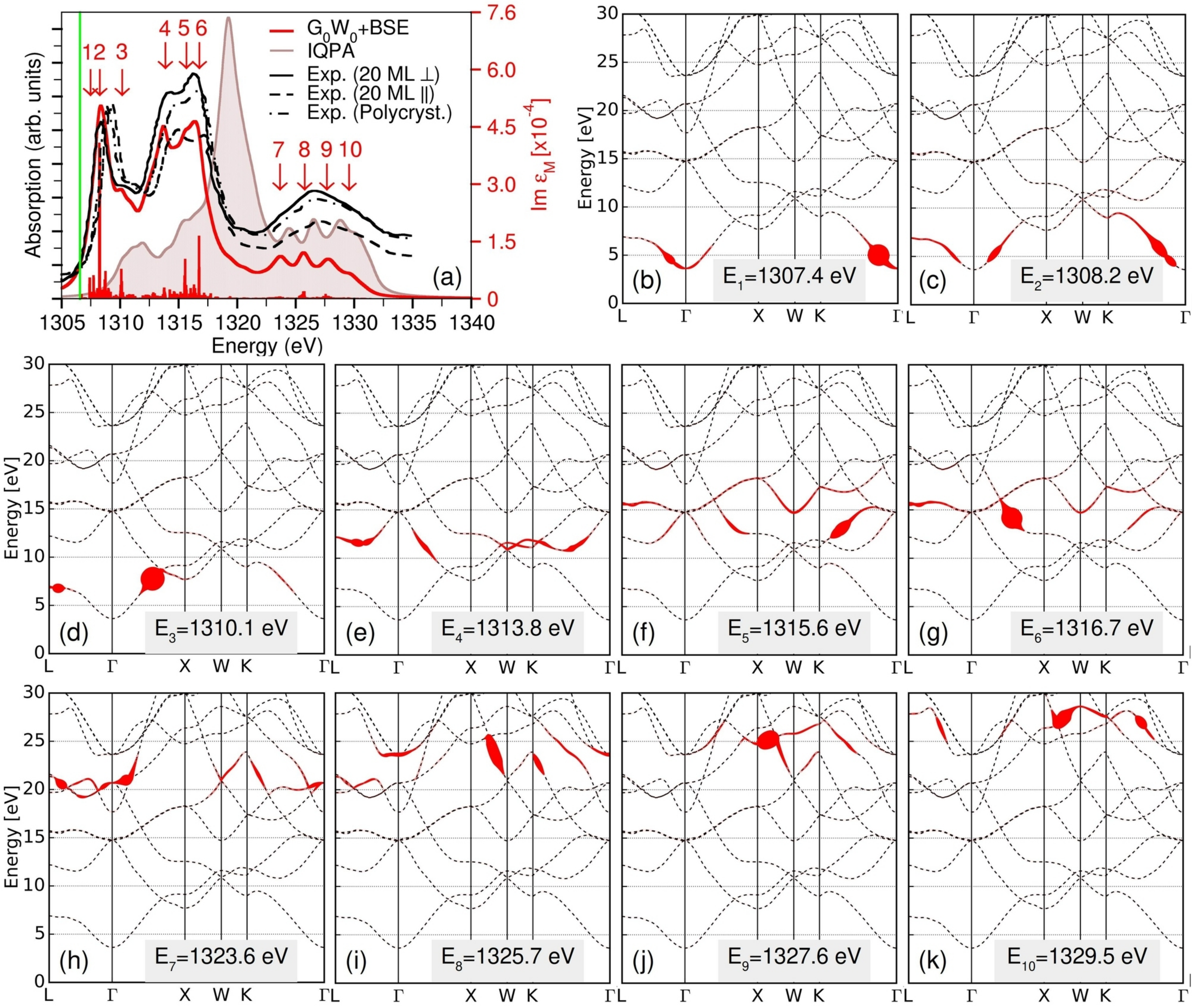}%
\caption{\label{fig:xas-Mg} XAS spectrum of Mg K-edge including $G_{0}W_{0}$+BSE
  corrections calculated with the \texttt{exciting} code: (a) calculated
  absorption spectra with $G_{0}W_{0}$+BSE (red line)) and within IQPA (brown shaded area)
  are compared with experimental spectra from Luches \textit{et al.}
  \cite{Luches} on a MgO film on Ag(001), grazing/normal incidence of photon beam
  (black dashed-dotted/solid line); polycrystalline MgO (black dashed line). A
  shift of 58.8 eV was applied to the calculated spectra to achieve coincidence
  with the first peak of the experiment and a Lorentzian broadening of 0.55 eV
  is adopted for the theoretical curve to mimic the excitation lifetime. The vertical green line at 1306.6 eV marks the direct band gap. The red vertical bars represent the oscillator strength (arb. units). (b-k)
  excitonic contributions to the final states in the CB of the peaks marked in (a).}
\end{figure*}

\subsubsection{\label{subsubsec:XAS-O} O K-edge}
The theoretical XAS spectrum of the O K-edge is plotted in Fig. \ref{fig:xas-O}d
together with the experimental spectrum from Luches \textit{et al.}
\cite{Luches}, who performed x-ray absorption measurements on MgO films of
varying thickness grown epitaxially on Ag(001) as well as on polycrystalline
bulk samples. The $G_0W_0+$BSE spectrum is characterized by six prominent peaks
with high oscillator strength (cf. Fig. \ref{fig:xas-O}a). Their origin in terms
of transitions to the conduction bands is visualized in Fig. \ref{fig:xas-O}b-g. 
We find that for the BSE calculation, a total of eight unoccupied bands are sufficient to obtain 
agreement with experiment and converge the oscillator strengths in the energy interval up to 30 eV.

While the spectrum within the independent quasiparticle aprroximation (IQPA) obtained after the 
$G_0W_0$ correction captures the  four peak-feature, a very good
agreement to experiment concerning the spectral shape and the relative positions of the three prominent peaks at
$\sim$ 537, 546, and 557 eV is obtained only after the $G_0W_0$+BSE
corrections. The spectrum is also consistent with earlier
work of Rehr \textit{et al.} \cite{Rehr} using FSR and BSE. The reduced intensity of the third peak in
the $G_0W_0$+BSE spectrum can be attributed to the limited number of unoccupied bands considered in the calculation.
Typically, the $GW$ approximation is not widely explored for core-level states, only recently there are 
first promising reports for its application to molecular $1s$ levels~\cite{Golze-2020}. In our calculations 
we do not correct the core energies obtained from DFT. Thus we shift the BSE spectrum to align 
with the experiment, as done in earlier works \cite{Vorwerk-2017,Vorwerk-2019}. For the O K-edge a shift of 
34.4 eV is applied to the $G_0W_0$+BSE spectrum to align the first peak with experiment. The same shift is 
also applied to the IQPA spectrum. 

The green line in Fig. \ref{fig:xas-O} a marks the direct band gap and the states below it correspond 
to bound excitons.The first bound exciton at the O K-edge is found at 534.5 eV with a binding energy of
690 meV, its real-space distribution is discussed in Section \ref{sec:RealspaceProj}. 
This value is comparable to previous results for other oxides, 
e.g. 0.5 eV was reported for $\beta$-Ga$_{2}$O$_{3}$~\cite{Cocchi-2016}, 
and 285 meV, 345 meV, and 323 meV for the $\alpha$-, $\beta$-, and $\varepsilon$ phase of 
Ga$_{2}$O$_{3}$~\cite{Vorwerk-2021}, respectively. 
Six prominent features in the XAS spectrum with high oscillator strength are 
marked and their origin is analyzed further in Figs. \ref{fig:xas-O}b-g. 
Transitions at the onset of the spectrum at 535.3 eV are localized around $\Gamma$ at the
bottom of the CB (Fig. \ref{fig:xas-O}b) and comprise predominantly O $3p$ character hybridized with 
O $3s$ (cf. Figs. \ref{fig:Vasp-Orbital}a-c), and Mg $3s$ character (cf. Fig.
\ref{fig:Vasp-Orbital}d). The second peak at 536.9 eV also arises from transitions to the lowest CB, 
but is more dispersive along $L-\Gamma-X$ and $K-\Gamma$. The subsequent peak at 537.7
eV stems from transitions to the lowest CB, but is localized midway along the
$L-\Gamma$ with some contribution along $K-\Gamma$ and has hybridized O $3s$, $3p$, 
and Mg $3s$ character. Furthermore, the peak at 539.5 eV results from
transitions to the second lowest unoccupied band localized at $X$ and dispersive
along $K-\Gamma$ with Mg 3$d_{xz}$ (cf. Fig. \ref{fig:Vasp-Orbital}f) as well as
O $3p$ character. Transitions to higher unoccupied bands around $W$ and $\Gamma$
with mixed O $3p$ and $3d$ and Mg $3p$, $t_{2g}$ character result in a peak at
546.5 eV. The final peak at 557.2 eV arises from transitions to CB at energies
above 25 eV with O $3p$ and Mg $3p$, $e_g$ contributions along $X-W-K-\Gamma$.

\begin{figure*}[!htp]
\includegraphics[width=0.7\textwidth]{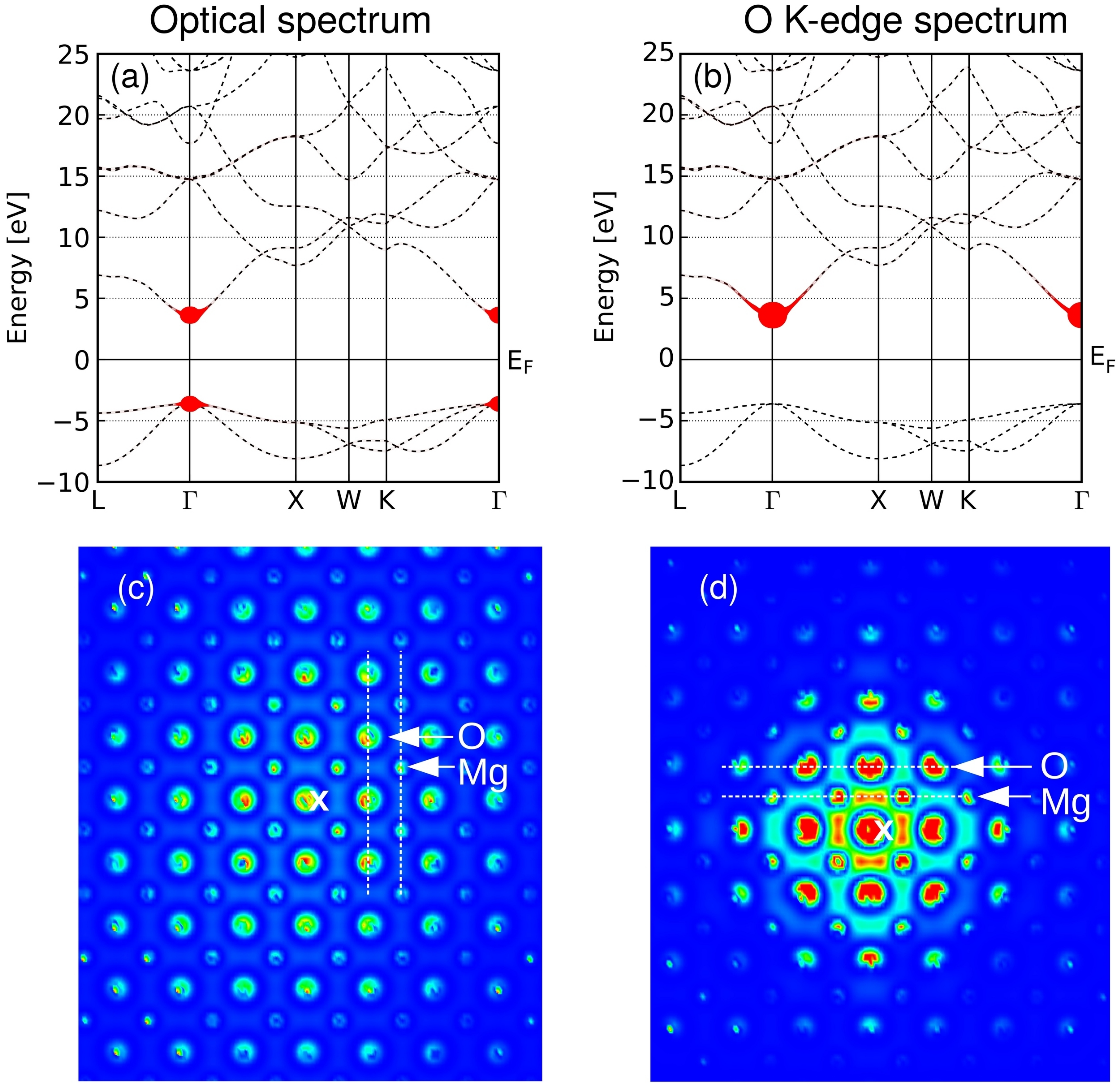}%
\caption{\label{fig:Exciton_O}Analysis of the first exciton in the optical spectrum (a) 
and O K-edge XAS (b) in reciprocal space. The lower panels show the density associated 
with the electronic part of the excitonic wavefunctions for a selected cross section in 
real space: In (c) along ($56\bar{1}$) for the optical excitation shown in (a) and 
(d) along ($\bar{5}1\bar{6}$) for the O K-edge XAS in (b). The color code visualizes 
the spacial extension of the wave functions: Blue colors refer to vanishing or 
low densities, orange to red colors to elevated densities. The hole is fixed near the oxygen 
(fractional coordinate: 0.52, 0.52, 0.52) and is marked by a white cross.}
\end{figure*}
\subsubsection{\label{subsubsec:XAS-Mg} Mg K-edge}
Fig. \ref{fig:xas-Mg}a displays the Mg K-edge from the $G_0W_0+$BSE calculation
and the experimental spectrum from Luches \textit{et al.} \cite{Luches}. The
experimental spectrum has four prominent peaks at 1308.3, 1314.4, and 1316.2 eV,
followed by a broader peak at 1326.6 eV, with a noticeable difference in
peak intensities for normal and grazing incidence of the MgO film on Ag(001) and for
the polycrystalline sample. While the IQPA spectrum for the O K-edge shows
overall agreement with the $G_0W_0$+BSE result, for the Mg K-edge the
IQPA spectrum fails to describe the general features of the experimental spectrum.
On the other hand, including the core-hole - electron interaction leads to a
large redistribution of the spectral weight, accompanied by the emergence of a
high intensity peak at the onset of the BSE spectrum. Overall, the $G_0W_0$+BSE
Mg K-edge is in very good agreement with the experimental spectrum of \citep{Luches}
and with previous BSE and FSR calculations \citep{Rehr} concerning peak
positions and relative intensity.
Similar to the O K-edge, we applied a shift of 58.8 eV to the $G_0W_0$+BSE spectrum to 
align it to the first peak in the experimental spectrum and the same shift is applied to the IQPA spectrum.

The green line in Fig.~\ref{fig:xas-Mg}a denotes the presence of bound excitons below this energy, 
the first bound exciton being at 1305.81 eV with a binding energy of 760 meV. 
Ten prominent peaks with high oscillator strength are marked in the spectrum which are analyzed
further in Figs. \ref{fig:xas-Mg}b-k. The first peak at 1307.4 eV arises from 
transitions to the bottom of CB with Mg $3p$ character, hybridized with Mg $3s$ states (cf.
Figs.~\ref{fig:Vasp-Orbital}a,b). The second peak at 1308.2 eV comprises 
transitions along the $L-\Gamma-X$ with Mg $3p$ character and along $K-\Gamma$
with mixed Mg $3s$ and $3p$ character. The third peak at 1310.1 eV includes
transitions to the lowest CB concentrated halfway between $\Gamma-X$ with
hybridized Mg $3s$ and O $3s$ and $3p$ character (cf. Figs.~
\ref{fig:Vasp-Orbital}a,b). Moreover, the peaks at 1313.8, 1315.6, and 1316.7 eV arise from transitions to 
higher energy CB ($>$10 eV) and are dispersive along the whole k-path with
hybridized O $3p$ and Mg $3s$ and $3p$ as well as Mg $3d_{xz}$ character (cf.
Figs.~\ref{fig:Vasp-Orbital}d-f). The peaks at 1323.6, 1325.7, 1327.6, and 1329.5
eV stem from the transitions to the unoccupied bands with $E>$20 eV
predominantly along $X-W-K$ with prevailing hybridized Mg $3p$ and $e_g$
character.

\subsection{\label{sec:RealspaceProj}Real-space projection of the first exciton}

The real-space wavefunction of the excited electron for a
given exciton can be obtained from the BSE
eigenvectors A$^\lambda$ as:
\begin{equation}\label{eqn:BSE_e-hWF}
\Psi_{\lambda}(\textbf{r}_{h},\textbf{r}_{e})=\sum_{vc\textbf{k}}A^{\lambda}_{vc\textbf{k}}\psi^{\ast}_{v\textbf{k}}(\textbf{r}_{h})\psi_{c\textbf{k}}(\textbf{r}_{e}).
\end{equation}
For more details see Ref. \cite{Sagmeister-2009,Vorwerk-2019} and references therein. For the analysis we fixed the hole slightly off the oxygen position (0.52,0.52,0.52) and plotted the electronic
part of the wavefunction in real-space for the first exciton of the optical and the O K-edge XAS spectrum in Fig.~\ref{fig:Exciton_O}.

The first bound exciton of the optical spectrum (Fig. \ref{fig:Exciton_O}a) consists of transition from the valence band maximum (VBM) to
the conduction band minimum (CBM) that are strongly localized around $\Gamma$.
Since the excited electron is distributed solely over the lowest, highly
dispersive conduction band, the bound exciton was previously described in the
Wannier-Mott two-band model by Fuchs \textit{et al.}~\cite{Fuchs-2008}. In Fig.~\ref{fig:Exciton_O}c we display a cut along the ($56\bar{1}$) plane through the center of the spread of the wave function near the fixed position of the hole, that shows that the exciton is delocalized  over several unit cells which supports
the Wannier-Mott character. Moreover the intensity of the spread has a maximum
at the oxygen sites and is weaker at the Mg sites. The reciprocal space projection in conjunction with the orbitally projected band structure (cf. Fig.~\ref{fig:Vasp-Orbital}) shows a main contribution of hybridized O $3s$ and Mg $3s$ states at the CBM.

For comparison, we have also analyzed the real space projection of the first
exciton in the O K-edge XAS spectrum. As shown in Fig. \ref{fig:Exciton_O}b this
exciton involves transitions to the CBM, but is more
dispersive in reciprocal space along $L-\Gamma-X$ and $K-\Gamma$. This goes hand
in hand with a stronger localization in real space, visible from the real space
projection in Fig. \ref{fig:Exciton_O}d along the ($\bar{5}1\bar{6}$) plane
that exhibits a significant decrease in the spread of the wavefunction. Compared
to the exciton of the optical spectrum, here the spread is confined to two to
three unit cells only. The 2D cut through the center of the
wavefunction spread also illustrates the orbital contributions with $s$ and $p$ character near the O sites, whereas the contributions around the Mg sites have $s$- like character. This can be attributed to the strong hybridization of the O $3s$, $3p$, and Mg $3s$ states around the CBM, discussed above.
\section{\label{sec:Summary}Summary}
We have provided a comprehensive study of the optical and x-ray absorption spectra of
bulk MgO with the VASP and \texttt{exciting} codes. The results indicate that
the quasiparticle, and in particular excitonic effects
are crucial to describe the spectra, concerning peak positions and to a lesser
extent intensity.

For the optical spectrum, the effect of two different functionals (GGA-PBEsol
and the hybrid HSE06) are studied: an excellent agreement with the experiment 
is obtained with HSE06 w.r.t. the energetic positions of the peaks. Analysis of
the electron-hole coupling coefficients in reciprocal space allows us to identify
the valence to conduction band transitions contributing to the peaks in the
spectrum. In particular, the peak at 7.3 eV arises due to transitions
localized around $\Gamma$ from the top of the VB to the bottom of the CB with mixed
O $3s$, $3p$, and Mg $3s$ character, followed by a peak at 9.4 eV stemming from
similar interband transitions but along $L-\Gamma-X$ and $\Gamma-K$ with mixed O
$3s$, $3p$, and Mg $3p$ character near $L$. The third peak at 10.4 eV is from
transitions to the bottom of CB from deeper lying valence bands and the final peak
at 12.2 eV results from a transition to higher lying conduction bands with hybridized  O $3p$ and Mg $3s$ and $3d_{xy}$,$d_{xz}$ character.

The inclusion of core-hole electron interaction by solving the BSE is found to
be essential also for the XAS Mg and O K-edge. By visualizing
the transitions to the unoccupied bands in reciprocal space,
we determine the origin of the relevant peaks in the spectra.
In the O K-edge spectrum, the peak at $\sim$ 537 eV originates from the
transitions to unoccupied bands with hybridized O $3s$, $3p$, and Mg $3s$
character, the peak at 546 eV stems from O $3p$, $3d$ hybridized with Mg $3p$
and $t_{2g}$ states, and the peak at 557 eV emerges from transitions to the CB
with hybridized O $3p$ and Mg $3p$ and 3d character. The real space projection 
of the electronic part of the wavefunction
of the first exciton  in the optical spectrum shows it has a delocalized
Wannier-Mott character, consistent with previous studies in reciprocal space
\cite{Fuchs-2008}. On the other hand, the wavefunction of the first exciton in the O K-edge 
spectrum is stronger localized and spreads up to only three unit cells. We believe that our 
detailed analysis of the optical and x-ray excitations in this paradigmatic oxide material 
regarding their orbital character and extension in real and reciprocal space based on 
state-of-the-art many-body approaches serves as an important benchmark and provides 
useful background information for the interpretation of experimental data both from 
static but also time-dependent investigations.

\begin{acknowledgments}
  We thank Caterina Cocchi, Andr\'e Schleife, Heiko Wende, Andrea Eschenlohr,
  Katharina Ollefs, Nico Rothenbach, and Okan K\"oksal for fruitful discussions.
  We wish to acknowledge funding by the Deutsche Forschungsgemeinschaft (DFG,
  German Research Foundation) within collaborative research center CRC1242
  (project number 278162697, subproject C02) and computational time at the
  Center for Computational Sciences and Simulation of the University of
  Duisburg-Essen on the supercomputer magnitUDE (DFG grants INST 20876/209-1
  FUGG, INST 20876/243-1 FUGG). C. V. and C. D. appreciate funding from the 
  Leibniz-ScienceCampus GraFOx. 
\end{acknowledgments}

\appendix

\begin{figure}[!htp]
\includegraphics[width=0.46\textwidth]{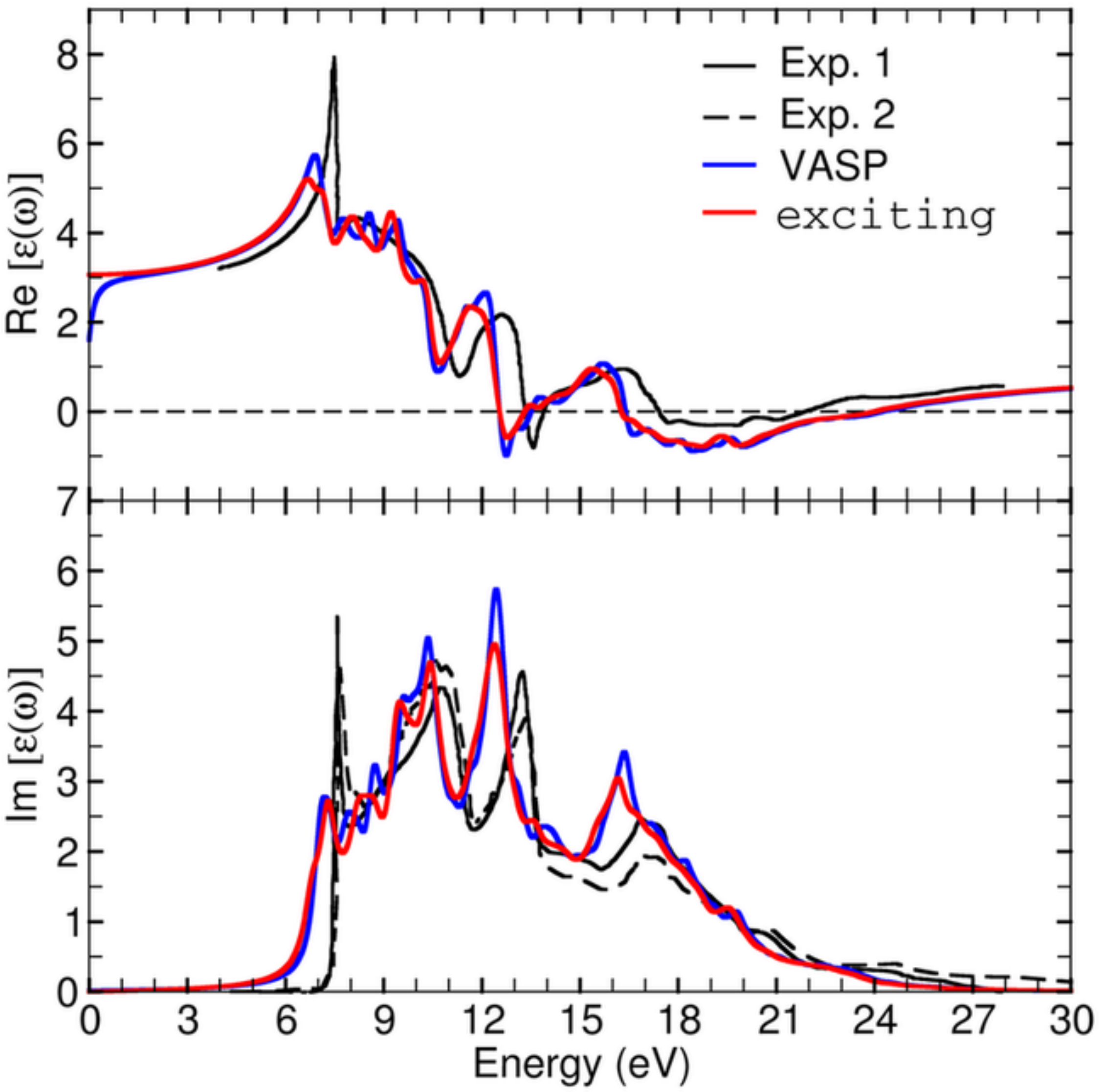}%
\caption{\label{fig:Comparison-OptSpec} Comparison of the $G_{0}W_{0}$+BSE spectrum calculated with PBEsol as the starting functional with \texttt{exciting} (red solid line) and VASP (blue solid line). The spectra are compared with the experiments of Roessler \textit{et al.} \cite{Roessler} (black solid line) and Bortz \textit{et al.} \cite{Bortz} (black dashed line). A Lorentzian broadening of 0.3 eV is employed for the theory spectra.}
\end{figure}
\section{\label{sec:Comp-OptSpec}Comparison of the optical spectrum calculated with \texttt{exciting} and VASP}

The optical spectra obtained with \texttt{exciting} and VASP with the same 
exchange-correlation functional (PBEsol) in Fig.~\ref{fig:Comparison-OptSpec} 
show very good agreement in the overall shape and peak positions with some 
smaller differences in peak heights.

\section{\label{sec:mBSE}DFT+model-BSE (mBSE)}

\begin{figure}[!htp]
\includegraphics[width=0.46\textwidth]{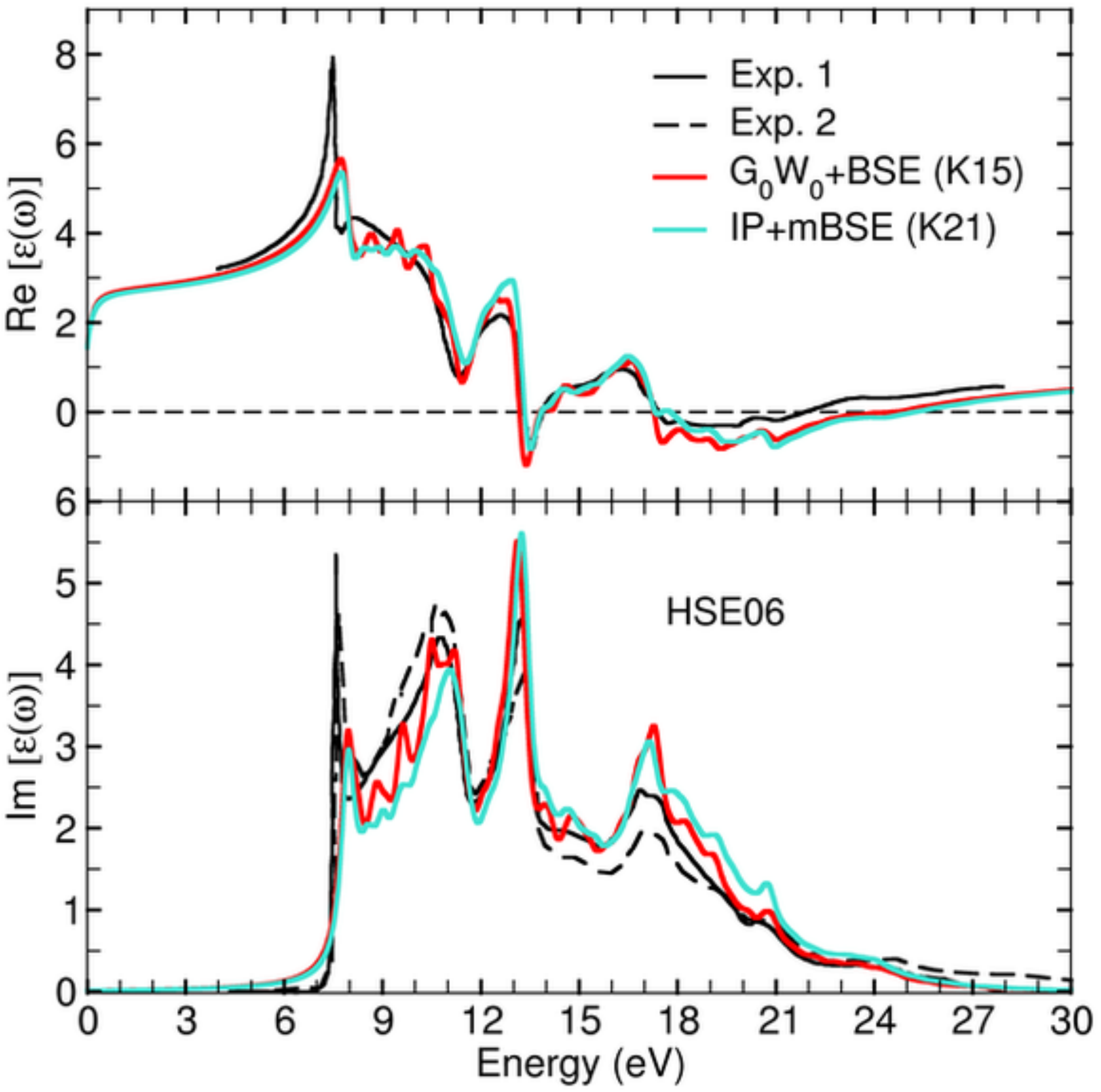}%
\caption{\label{fig:optics-mBSE} Comparison of the spectrum obtained with mBSE
  on top of DFT (\textbf{k}-mesh of 21$\times$21$\times$21) and with
  $G_{0}W_{0}$+BSE corrections (\textbf{k}-mesh of 15$\times$15$\times$15). The
  spectra were calculated with HSE06 as the starting functional and compared
  with the experiments of Roessler \textit{et al.} \cite{Roessler} (black solid
  line) and Bortz \textit{et al.} \cite{Bortz} (black dashed line). A Lorentzian
  broadening of 0.3 eV is employed for the theory spectra.}
\end{figure}

Fuchs \textit{et al.} \cite{Fuchs-2008} have pointed out the importance of using
a dense $k$-mesh for BSE in order to obtain convergence of the
optical spectrum in the lower energy range. However, this goes hand in hand with
a high computation cost in the $G_{0}W_{0}$ calculation. One way to circumvent
this is to perform a model BSE calculation directly using the DFT wave
functions and considering only the required number of bands which cover the
energy range of interest \cite{Fuchs-2008}. The omission of the $G_{0}W_{0}$
step allows us to use a higher $k$-mesh and thus improve the convergence. 
Here, we discuss the result obtained by using a model for the static screening 
with parameters fitted to the screened Coulomb kernel diagonal values obtained 
from $G_{0}W_{0}$ calculation, a detailed description can be found in 
Ref.~\cite{Fuchs-2008,Bokdam,Bechstedt,Liu} and was previously used in 
Ref.~\cite{Bokdam,Liu}. Good agreement between mBSE and the full BSE
($G_{0}W_{0}+$BSE) was recently obtained for bulk SrTiO$_3$~\cite{Begum}.

In our study, mBSE is performed on a 21$\times$21$\times$21 $k$-mesh with a
range separation parameter $\lambda=1.44$ \AA$^{-1}$, ion-clamped static
dielectric function $\epsilon_{\infty}^{-1}=0.38974$, and a scissor shift of 1.95
eV, corresponding to the difference between the KS energies obtained with HSE06
and the quasiparticle band gap. In Fig. \ref{fig:optics-mBSE} the spectrum from
mBSE and full BSE obtained with HSE06 as the starting functional on a
\textbf{k}-mesh of 15$\times$15$\times$15 are displayed. Overall, very good
agreement is obtained. In particular, in the energy range of 7 $-$ 11
eV we even observe a better agreement with experiment for mBSE when compared
with full BSE associated with the increase in $k$-point density. For energies
higher than 11 eV, the positions of the peaks are also well reproduced with the
mBSE approach with slight increase in the intensity when compared with the full
BSE spectrum. The good agreement between the mBSE and full BSE spectra is
attributed to the fact that the electronic structure does not change
significantly from DFT to $G_{0}W_{0}$ (cf. Fig. \ref{fig:Bandstr-PDOS}), the
prevailing effect being a rigid shift of the conduction band. The binding energy
of the first exciton from mBSE is 610.8 meV.


\providecommand{\noopsort}[1]{}\providecommand{\singleletter}[1]{#1}%

\end{document}